\documentclass[12pt,russian]{iopart}
\usepackage{iopams}
\usepackage{epsfig}
\usepackage[english,russian]{babel}
\begin{document}


\title{Hybrid Taylor-WKB series}

\author{N L Chuprikov}

\address{Tomsk State Pedagogical University, 634041, Tomsk, Russia}
\begin{abstract}
A generalized WKB approach for constructing WKB series endowed with some properties
of Taylor ones is presented. Apart from the Riccati equation itself its formalism
involves also the Riccati-equation's derivatives (REDs) obtained by differentiating
of the former with respect to a spatial variable. For any smooth potential barrier
given in the finite spatial interval to include turning points, the zeroth-order
term of presented WKB series is regular everywhere in this interval. Moreover, the
more REDs are used, the more exact the zeroth-order solution is.
\end{abstract}

\maketitle

\newcommand{\ppp}{\mbox{\hspace{5mm}}}
\newcommand{\ooo}{\mbox{\hspace{3mm}}}
\newcommand{\ooa}{\mbox{\hspace{1mm}}}

\newcommand{\xx}{(x;\hbar)}
\newcommand{\ee}{\epsilon\xx}

\section{Introduction} \label{a20}

In 1994 year two generalized WKB approaches has been proposed - in the paper
\cite{Doo}, for solving the Helmholtz equation, and in the papers \cite{Ch10,Ch11}
(see also \cite{Ch12}), for solving the one-dimensional Schr\"odinger equation
(OSE). Both are based on the same idea - to use the REDs in order to obtain WKB
series regular at the turning points (later, in 2002 year, this idea has been
reopened in \cite{Adac}). However, despite the common idea, these approaches differ
from each other, because they treat the REDs in a different way. For example, in the
approach \cite{Doo} the searched-for solutions are expanded traditionally in integer
powers of Plank's constant $\hbar$. While in the formalism \cite{Ch10,Ch11} the role
of a small parameter is played by some function $\ee$ of the spatial variable $x$
and Plank's constant. In this paper we develop the approach \cite{Ch10,Ch11,Ch12}.

\section{Formalism} \label{a22} Let us consider the OSE
\begin{equation} \label{201}
\fl \frac{d^2\Psi}{dx^2}-Q^{(0)}\xx\Psi=0, \ppp Q^{(0)}\xx=
\frac{2m_e}{\hbar^2}\left(V(x)-E\right);
\end{equation}
where the potential $V(x)$ is zero beyond the interval $[a,b]$ and is a bounded
infinitely differentiable function; $E$ is the energy of a particle; $m_e$ is its
mass. It is also assumed that there are turning points in the interval $[a,b]$ whose
order does not exceed $N$. Remind, if $x_0$ is the turning point of the $n$-th order
(here $0\leq n\leq N$), then
\begin{equation} \label{2000}
\fl Q^{(0)}(x_0)=\ldots=Q^{(n-1)}(x_0)=0; \ooo Q^{(n)}(x_0)\ne 0;
\end{equation}
$Q^{(k)}=d{[Q^{(0)}]}^k/dx^k$; if $n=0$, points $x_0$ can be formally considered as
zeroth-order turning points.

Letting, as in the standard WKB approach, $\Psi(x)=e^{S(x)}$ and $y(x)=dS(x)/dx$, we
reduce solving Eq.~(\ref{201}) to solving the Riccati equation. However, unlike the
standard WKB approach, we write down this equation together with the first $N$ REDs:
\begin{eqnarray} \label{204}
\fl y^{(m+1)}+\sum^m_{k=0}C^m_k y^{(m-k)}y^{(k)}-Q^{(m)}=0; \ooo
\frac{dy^{(N)}}{dx} +\sum^N_{k=0} C^N_k y^{(N-k)}y^{(k)}-Q^{(N)}=0;
\end{eqnarray}
$m=0,\ldots,N-1$; $y^{(k)}=dy^k/dx^k$;  $C^m_k$ are the binomial coefficients. The
derivatives $y^{(m)}\xx$ ($m=1,\ldots,N$) are considered here as independent
functions to obey the set of Eqs.~(\ref{204}); the equality
$y^{(m+1)}\xx=dy^{(m)}\xx/dx$ holds, provided that Eqs. (\ref{204}) are solved
exactly.

Let now
\begin{eqnarray} \label{2040}
\fl y^{(m)}\xx=\epsilon^{-m-1}\xx u_m\xx \hspace{8mm}(m=0,\ldots,N),
\end{eqnarray}
where $\ee$ is some function whose norm $\|\ee\|$ diminishes when $\hbar\to 0$;
thereinafter the norm $\|f(x)\|$ of any function $f(x)$ is determined as the maximum
value of $|f|$ in the interval $[a,b]$. Substitution of Exp.~(\ref{2040}) into
Eq.~(\ref{204}) yields
\begin{eqnarray} \label{205}
\fl u_{m+1}+\sum^m_{k=0}C^m_k u_{m-k}u_{k}-Q_{m}=0; \ooo \epsilon
u'_N-(N+1)\epsilon'u_N+\sum^N_{k=0} C^N_k u_{N-k}u_{k}-Q_{N}=0
\end{eqnarray}
where $m=0,\ldots,N-1$; $Q_\ell=\epsilon^{\ell+2}Q^{(\ell)}$; $\ell=0,\ldots,N$; the
primes denote the differentiation with respect to $x$.

Now let us search-for the solution of Eqs.~(\ref{205}) in the form of the expansion
in integer powers of $\ee$,
\begin{equation} \label{206}
\fl u_n\xx=\sum_{k=0}u_{m,k}\xx \epsilon^k\xx,
\end{equation}
supposing that for any $k$ and $m$ the following relations are valid
\begin{eqnarray} \label{207}
\fl \left\|\frac{d^k\ee}{dx^k}\right\|\sim\left\|\ee\right\|,
\ooo\left\|Q_k\xx\right\|=O(1),\ooo  \left\|u_{m,k}\xx\right\|=O(1).
\end{eqnarray}

Keeping the zeroth-order terms, we obtain
\begin{eqnarray}\label{208}
\fl u_{m+1,0}+\sum^m_{k=0}C^m_k u_{m-k,0}u_{k,0}=Q_{m} \ooo (m=0,\ldots,N-1); \ooo
\sum^N_{k=0} C^N_k u_{N-k,0}u_{k,0}=Q_{N}
\end{eqnarray}
(we have to stress once more that the functions $Q_0, \ldots, Q_N$ are of the same
order, by norm; it should be taken into account that the function $|Q_N(x;\hbar)|$
takes its maximal value at the turning point (of the $N$-th order), however the ones
$|Q_0(x;\hbar)|, \ldots, |Q_{N-1}(x;\hbar)|$ approach their maxima beyond this
point). In the $\alpha$-th order ($\alpha=1,2,\ldots$),
\begin{eqnarray} \label{209}
\fl u_{m+1,\alpha}+\sum^m_{k=0}\sum^\alpha_{\gamma=0}C^m_k
u_{m-k,\gamma}u_{k,\alpha-\gamma}=0 \ooo \ooo (m=0,\ldots,N-1); \nonumber\\
\fl \frac{du_{N,\alpha-1}}{dx}-(N+2-\alpha)\frac{\epsilon'}{\epsilon}
u_{N,\alpha-1}+ \sum^N_{k=0} C^N_k\sum^\alpha_{\gamma=0}
u_{N-k,\gamma}u_{k,\alpha-\gamma}=0.
\end{eqnarray}

Now we have to define the unknown function $\ee$ to enter Eqs.~(\ref{208}) via the
functions $Q_m\xx$ ($m=0,\ldots,N$). For this purpose we will suppose that for all
$m$ $u_{m,0}=m!a_m$ where $a_0=1$. Besides, for convenience let further
$\kappa=\epsilon^{-1}$. Then, from Eqs. (\ref{208}), we obtain
\begin{eqnarray} \label{2011}
\fl (m+1)a_{m+1}+\sum^m_{k=0}a_{m-k}a_{k}=\frac{Q^{(m)}}{m!\ooa \kappa^{m+2}} \ppp
(m=0,\ldots,N-1); \nonumber\\\fl \sum^N_{k=0} a_{N-k}a_k=\frac{Q^{(N)}}{N!\ooa
\kappa^{N+2}}.
\end{eqnarray}
These equations can be reduced to the algebraic equation of the ($N+2$)-th order for
$\kappa\xx$. For example, for $N=0$, $N=1$ and $N=2$ we have, respectively,
\begin{eqnarray} \label{2019}
\fl \kappa^2=Q^{(0)},\ooo\kappa^3-\kappa Q^{(0)}+\frac{Q^{(1)}}{2}=0,\ooo
3\kappa^4-4\kappa^2 Q^{(0)}+\kappa Q^{(1)}+
\left(Q^{(0)}\right)^2-\frac{Q^{(2)}}{2}=0.
\end{eqnarray}

\section{The relationship between small parameters $\ee$ and $\hbar$ at the turning
point of the $n$-th order} \label{a23}
\newcommand{\ku}{\tilde{Q}}
Let us now show that there are at least two roots $\kappa$ of Eqs. (\ref{2011}),
such that
\begin{equation} \label{2013}
\fl \lim_{\hbar\to 0}|\kappa^{-1}\xx|=0
\end{equation}
for any point $x\in[a,b]$. For example, let $x$ be the turning point of the $n$-th
order; $\left|Q^{(n+1)}(x)/Q^{(n)}(x)\right|\ll|\kappa(x;\hbar)|$. The solution of
Eqs. (\ref{2011}), at this point, can be found in the form
\newcommand{\kk}{\tilde{\kappa}}
\newcommand{\ao}{\tilde{a}}
\begin{equation} \label{20140}
\fl \kappa(x;\hbar)=\hbar^{-\beta}\sum_{s=0}^\infty\kk_{s}(x)\hbar^{s\beta};\ooo
a_m(x;\hbar)=\sum_{s=0}^\infty a_{m,s}(x)\hbar^{s\beta}\ooo (m=1,\ldots,N)
\end{equation}
where $\beta$ is an arbitrary positive constant. The substitution of Exps.
(\ref{20140}) into Eqs. (\ref{2011}), with taking into account (\ref{2000}) and
keeping only the main-order terms, yields that the series (\ref{20140}) are
justified only for $\beta=2/(n+2)$. In this case
\begin{eqnarray} \label{2017}
\fl (m+1)a_{m+1,0}+\sum^m_{k=0}a_{m-k,0}a_{k,0}=0 \ooo (m=0,\ldots,n-1,n+1,\ldots,N-1);\nonumber\\
\fl \sum^N_{k=0} a_{N-k,0}a_{k,0}=0
\end{eqnarray}
where it is assumed that $a_0=a_{0,0}=1$. When Eqs. (\ref{2017}) have been solved
one can find $\kk_0$ from the equation
\begin{eqnarray} \label{2018}
\fl \left[(n+1)a_{n+1,0}+\sum^n_{k=0}
a_{n-k,0}a_{k,0}\right]\kk^{n+2}_0=\frac{2m_e}{n!}V^{(n)}(x_0).
\end{eqnarray}

From Eq. (\ref{2018}) it follows that there are $n+2$ complex roots $\kk_0$ with
equal absolute values at the turning point of the $n$-th order. This means that
there are $n+2$ different functions $\kappa(x;\hbar)$ such that at this point
\begin{eqnarray} \label{2033}
\fl |\kappa(x;\hbar)|\sim\hbar^{-2/(n+2)},
\end{eqnarray}
thereby the condition (\ref{2013}) is fulfilled for these roots. If $n=0$, i.e., if
$x$ is not a turning point, then there are two roots to satisfy (\ref{2013}).

So, the expansion in the integer powers of $\ee$ at the turning point of the $n$-th
order is equivalent to that in the small parameter $\hbar^{2/(n+2)}$. This is in a
full agreement with the existing approaches. Far from the turning points the
expansion (\ref{206}) is equivalent to that in integer powers of $\hbar$, as in the
WKB-approach.

What is important, one of two relevant roots $\kappa(x;\hbar)$ of Eqs. (\ref{2011})
obeys the condition (\ref{2013}) for any point $x\in[a,b]$ (hereinafter this root
will be denoted as $\kappa^{(1)}(x;\hbar)$). For other roots this condition breaks
when $x$ crosses turning points. Thus, only one root of these equations yields the
function $\ee$ to have a small norm in the interval $[a,b]$.

The zeroth-order solution, $y\approx [\ee]^{-1}=\kappa^{(1)}\xx$, associated with
this root yields a good approximation of the exact solution of the Riccati equation
in the whole interval $[a,b]$. In this case, the more the REDs are taken into
account, the more precise this approximation is. We do not give a strong proof of
this statement. Instead we pay reader's attention to the fact that for a smooth
potential, in the limit $N\to\infty$, the set of Eqs.~(\ref{2011}) for the zeroth
order solution coincides with the exact set of Eqs. (\ref{204}), considering that in
the zeroth order $y^{(m)}=m!a_m \kappa^{m+1}$.

As is seen, the presented WKB series like the Taylor ones includes the derivatives
of the function expanded; and the larger the number considered derivatives, the more
precise the zeroth-order WKB-approximation is. In this sense the presented
generalized WKB series can be considered as a hybrid Taylor-WKB series. A
distinctive feature of such series is the possibility to improve the exactness of
the zeroth-order WKB-approximation. Such possibility is very important, because
considering of nonzero terms in asymptotic approaches is usually associated with
serious mathematical problems (and our approach is not an exclusion; however, in our
approach, Eqs.~(\ref{209}) are extra in fact).

\section{Example: a linear potential} \label{a24}

To exemplify our approach, let us consider the case when an electron with the energy
$E=0.1eV$ impinges the linear potential $V(x)=(1+x/d)V_0$ to be nonzero in the
interval $[-d/2,d/2]$; $V_0=0.1$eV, $d=100nm$. In such setting the point $x=0$ is a
turning point of the first order.

\begin{figure}
\begin{center}
\includegraphics[width=8cm,angle=-90]{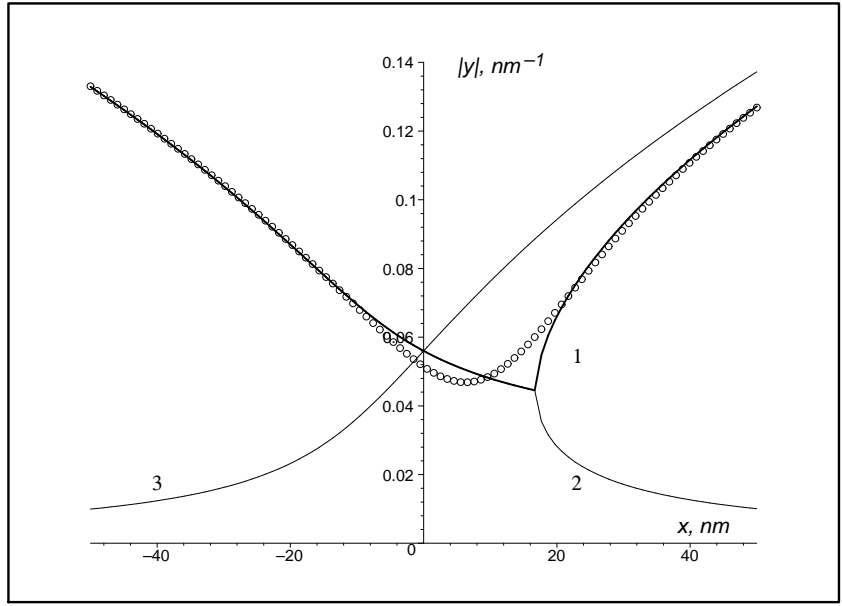}
\end{center}
\caption{Absolute values of the exact solution $y(x)$ (circles) and the all three
roots $\kappa(x;\hbar)$ of Eq.~(\ref{2019}) for $N=1$.} \label{fig:fig1}
\end{figure}

\begin{figure}
\begin{center}
\includegraphics[width=8cm,angle=-90]{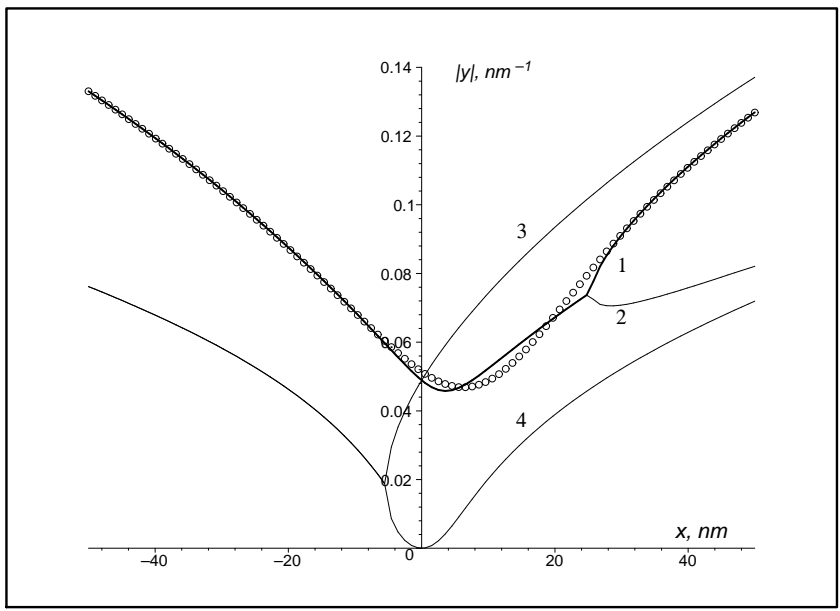}
\end{center}
\caption{Absolute values of the exact solution $y(x)$ (circles) and the all four
roots $\kappa(x;\hbar)$ of Eq.~(\ref{2019}) for $N=2$.} \label{fig:fig2}
\end{figure}

Figs. \ref{fig:fig1} and \ref{fig:fig2} show, respectively, the absolute values of
the solutions of the second and first equations (\ref{2019}) as well as those of the
exact solution of the Riccati equation
\begin{eqnarray} \label{20180}
\fl y(x)=\rho \frac{Bi^\prime(\rho x)+i\cdot Ai^\prime(\rho x)} {Bi(\rho x)+i\cdot
Ai(\rho x)}
\end{eqnarray}
where $Ai$ and $Bi$ are the Airy functions, à $Ai^\prime$ and $Bi^\prime$ are their
first derivatives, respectively; $i=\sqrt{-1}$, $\rho\approx 7,055\cdot 10^{-2}\ooa
nm^{-1}$; the solution is obtained with the help of the Maple programme. Solution
(\ref{20180}) is evident to correspond to the solution $\Psi(x)=Bi(\rho x)+i\cdot
Ai(\rho x)$ of Eq.~(\ref{201}).

Fig. \ref{fig:fig1} shows that in the limit $x\to\infty$ the condition (\ref{2013})
holds for the roots 1 and 3, while in the limit $x\to\-\infty$ it does for the roots
1 and 2 which are the complex conjugates to each other in this region. Thus,
$\kappa^{(1)}(x;\hbar)$ is the root 1 for $N=1$. For $N=2$ (see Fig. \ref{fig:fig2})
the same numeration is used for relevant roots.

The roots $\kappa^{(1)}(x;\hbar)$ of Eqs. (\ref{2019}) which give the best
approximation of $y(x)$ for $N=0$, $N=1$ and $N=2$ can be presented in an analytical
form as follows. For $N=0$ (the first equation in the set (\ref{2019})),
$\kappa^{(1)}(x;\hbar)=\left[(Q^{(0)}\xx)^{1/2}\right]^*$; hereinafter, $-\pi/n\leq
arg(z^{1/n})\leq \pi/n$ for any complex $z$ and integer $n$; besides, we take into
account here that if $z$ is a complex root of an algebraic equation with real
coefficients, then its complex conjugate $z^*$ is its root too.

For $N=1$ (the second equation in the set of Eqs. (\ref{2019}))
\begin{eqnarray} \label{300}
\fl \kappa^{(1)}(x;\hbar)=\frac{1}{2}(A_1+A_2)-i\frac{\sqrt{3}}{2}(A_1-A_2), \\
\fl  A_{1,2}=surd(r\pm\sqrt{q^3+r^2},3);\ooo q=-\frac{Q^{(0)}(x)}{3},\ooo
r=-\frac{Q^{(1)}(x)}{4};\nonumber
\end{eqnarray}
here $surd(z,3)$ is the Maple's function of a complex variable $z$: if $\Re(z)\geq
0$ then $surd(z,3)=z^{1/3}$; otherwise, $surd(z,3)=-(-z)^{1/3}$; note that when the
function $z^{1/3}$ is used straightforwardly in the Exps. (\ref{300}), this
analytical expression yields different roots in the different parts of the interval
$[a,b]$.

From Exps. (\ref{300}) it follows that far from the turning point $x=0$ (when the
term with $Q^{(1)}$, in the expression for $A_{1,2}$, is negligible)
$|\kappa^{(1)}(x;\hbar)|\sim\hbar^{-1}$; but at the turning point itself, where
$Q^{(0)}=0$, $|\kappa^{(1)}|=\sqrt[3]{|Q^{(1)}/2|}\sim\hbar^{-2/3}$ (see also
(\ref{2033})).

Lastly, for $N=2$ (the third equation in the set (\ref{2019})) we have
\begin{eqnarray} \label{301}
\fl \kappa^{(1)}(x;\hbar)=-u(x)+\sqrt{(u(x))^2-v(x)},\nonumber\\
\fl u= -\frac{1}{2}\sqrt{z(x)-a_2(x)},\ooo
v=\frac{z(x)}{2}+\sqrt{\left(\frac{z(x)}{2}\right)^2-a_0(x)},
\\\fl a_2= -\frac{4}{3}Q^{(0)}(x);\ooo a_1= \frac{1}{3}Q^{(1)}(x),
\ooo
a_0=\frac{1}{3}\left[\left(Q^{(0)}(x)\right)^2-\frac{1}{2}Q^{(2)}(x)\right];\nonumber
\end{eqnarray}
here $z$ is the solution of the auxiliary equation
\begin{eqnarray} \label{302}
\fl z^3+b_2(x)z^2+b_1(x)z+b_0(x)=0;
\end{eqnarray}
where $b_2=-a_2(x)$, $b_1=-4a_0(x)$, $b_0=4a_0(x)a_2(x)-(a_1(x))^2$. This solution
is defined as follows
\begin{eqnarray} \label{303}
\fl z=s_1(x)+s_2(x)-\frac{b_2(x)}{3},\ooo
s_{1,2}=\left[\nu(x)\pm\sqrt{\mu(x)^3+\nu(x)^2}\right]^{1/3},
\\\fl \mu=\frac{b_1(x)}{3}-\frac{b_2(x)^2}{9},
\ooo \nu=\frac{1}{6}\left[b_1(x)b_2(x)-3b_0(x)\right]-\frac{b_2(x)^3}{27}.\nonumber
\end{eqnarray}

\begin{figure}
\begin{center}
\includegraphics[width=8cm,angle=-90]{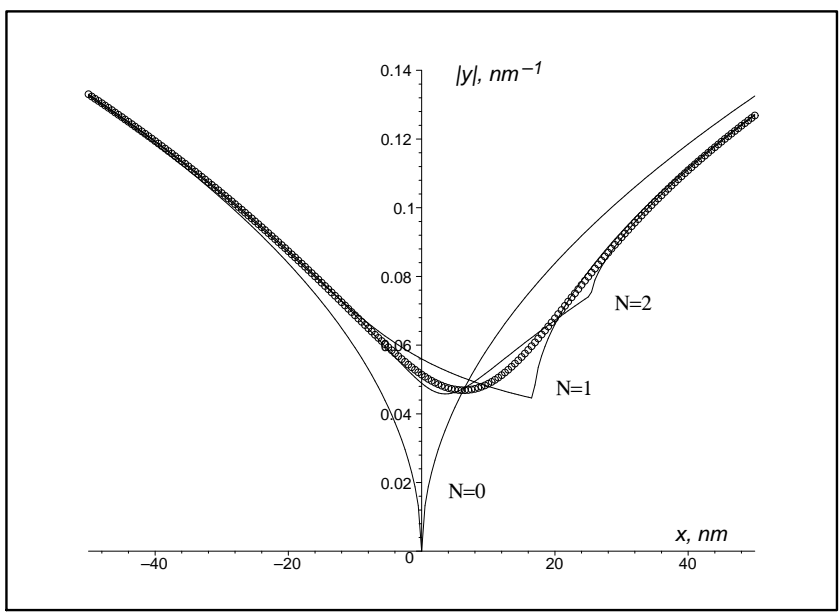}
\end{center}
\caption{Absolute values of the exact solution $y(x)$ (circles) and the roots
$\kappa^{(1)}(x;\hbar)$ for $N=0$, $N=1$ and $N=2$.} \label{fig:fig3}
\end{figure}

\begin{figure}
\begin{center}
\includegraphics[width=8cm,angle=-90]{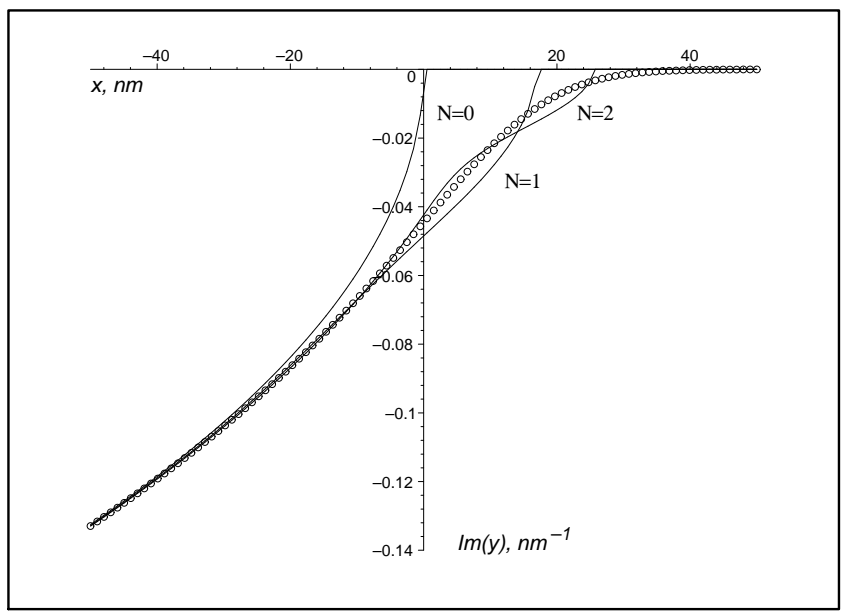}
\end{center}
\caption{Imaginary parts of the exact solution $y(x)$ (circles) and the roots
$\kappa^{(1)}(x;\hbar)$ for $N=0$, $N=1$ and $N=2$.} \label{fig:fig4}
\end{figure}

Figs.~\ref{fig:fig3} and \ref{fig:fig4} show the $x$-dependence of the absolute
values and imaginary parts, respectively, of $\kappa^{(1)}(x;\hbar)$ obtained for
$N=0$, $N=1$ and $N=2$. As is seen, the more the REDs are used in the formalism, the
better the zeroth-order approximation is.
\begin{figure}
\begin{center}
\includegraphics[width=8cm,angle=-90]{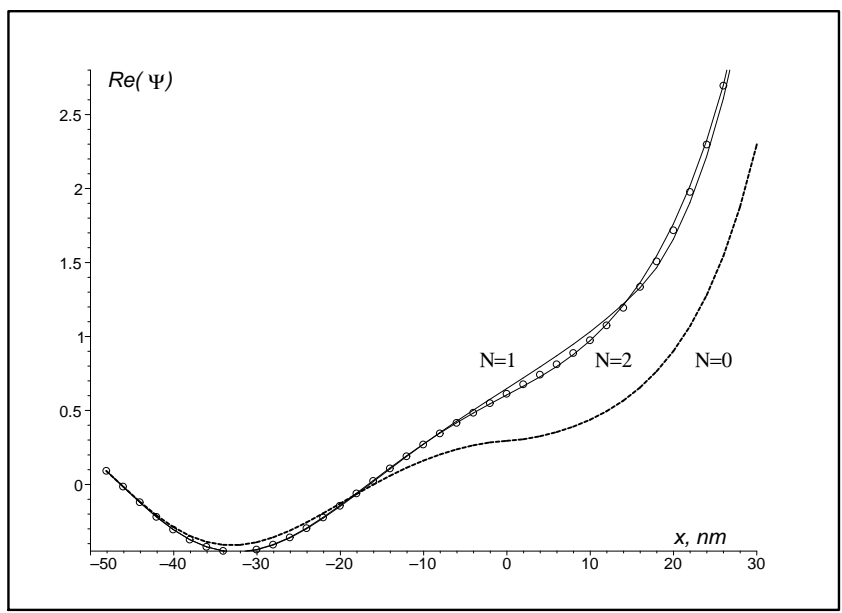}
\end{center}
\caption{Real parts of the exact wave function $\Psi(x)$ (circles) and its
zeroth-order approximations for $N=0$, $N=1$ and $N=2$.} \label{fig:fig5}
\end{figure}
\begin{figure}
\begin{center}
\includegraphics[width=8cm,angle=-90]{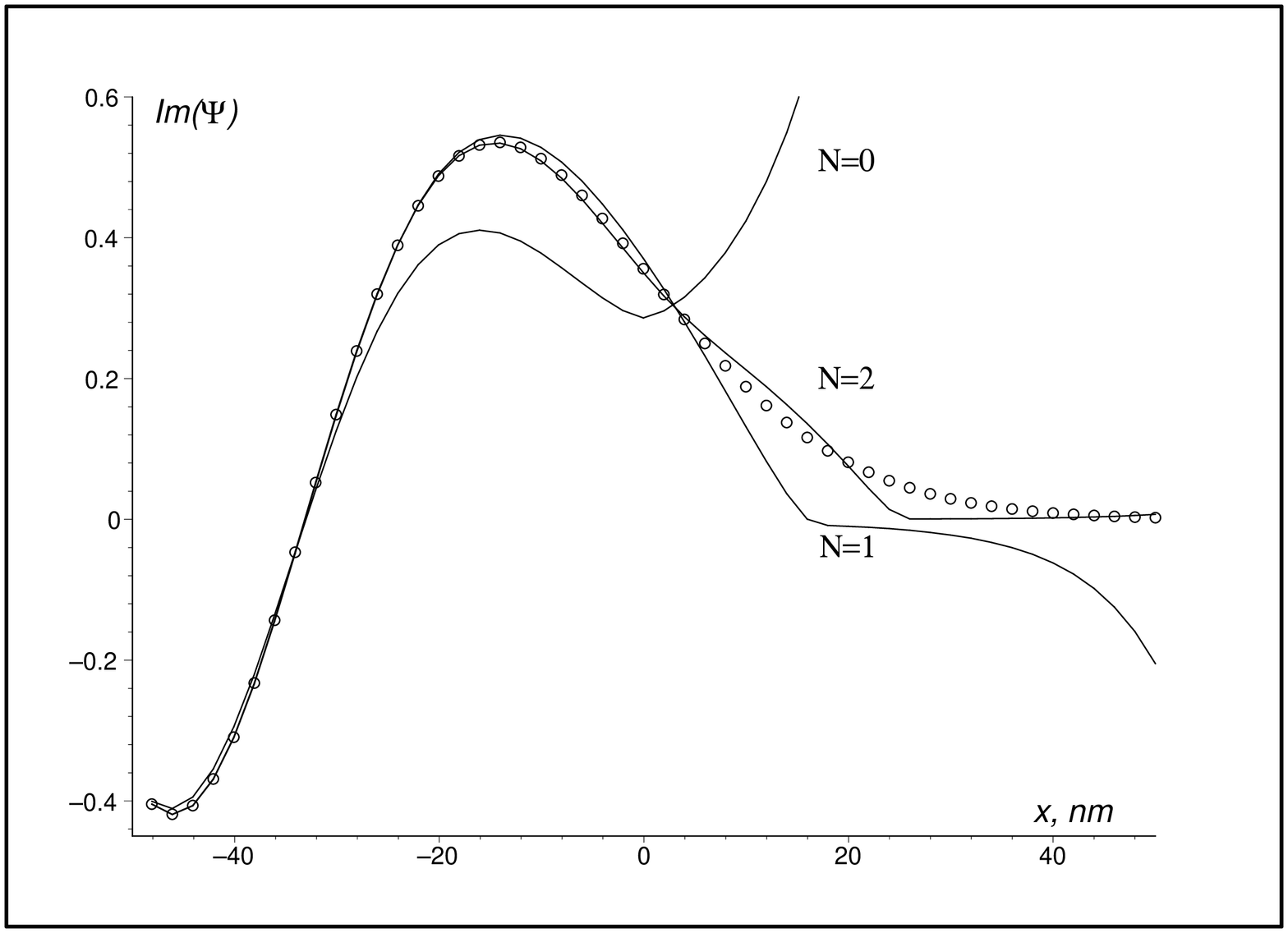}
\end{center}
\caption{Imaginary parts of the exact wave function $\Psi(x)$ (circles) and its
zeroth-order approximations for $N=0$, $N=1$ and $N=2$.} \label{fig:fig6}
\end{figure}

We have to stress that for the potential at hand and for any $N$ there is always
such a point $x_b$ in the region $x>0$ (the below-barrier region) that
$\Im\left(\kappa^{(1)}(x;\hbar)\right)=0$ for $x\geq x_b$ (see Fig.~\ref{fig:fig4}).
At the same time for the exact solution $y(x)$ considered here there are no regions
on the $OX$-axis where $\Im(y(x))\equiv 0$. Though $|\Im(y(x))|\ll|\Re(y(x))|$ in
the below barrier region, $\Im(y(x))$ cannot be neglected in any case, because it
yields the second independent real solution to the OSE, i.e., the function $Ai(\rho
x)$. As regards the approximate solution $\kappa^{(1)}(x;\hbar)$, if $x_b\in[a,b]$
then in the region $[x_b,b]$ it loses this solution. However, as is seen from
Fig.~\ref{fig:fig4}, this point shifts to the right on the OX axis when $N$
increases. Thus, to eliminate the above shortcoming, one has to include the next
RED's into the formalism of the zeroth-order approximation. For a given potential
$V(x)$ and given spatial interval $[a,b]$, there exist such a value of $N$ for which
$x_b>b$. This property exemplifies the above statement that the zeroth-order
approximation $\kappa^{(1)}(x;\hbar)$ becomes precise in the limit $N\to\infty$.

On Figs.~\ref{fig:fig5} and \ref{fig:fig6} we show, respectively, the real and
imaginary parts of the exact solution $Bi(\rho x)+i Ai(\rho x)$ of the OSE and its
zeroth-order generalized WKB-approximations for $N=0$, $N=1$ and $N=2$. As is seen,
even for the exponentially decaying (for $x>0$) function Airy $Ai(\rho x)$ the
zeroth-order approximation for $N=2$ gives a good exactness. Moreover, the real part
of this approximation fits perfectly the Airy function $Bi(\rho x)$. As regards the
zeroth-order approximation at $N=0$, for this case there is no root
$\kappa^{(1)}(x;\hbar)$ which would fit both the independent real solutions of the
OSE in the below-barrier region ($x>0$).

\section{Conclusion}

So, for constructing everywhere regular WKB-series it is suggested to include into
the WKB formalism the spatial derivatives of a potential-energy function under
study. WKB-series obtained in such a way are named here the Taylor-WKB ones. For a
smooth potential given in the finite spatial interval the zeroth-order term of the
Taylor-WKB series yields a good approximation in the whole interval under study,
including turning points if they exist. The more the derivatives of the potential
are involved into the formalism, the more exact the zeroth-order approximation is.
Of importance is the fact that for smooth potentials to have only the first- and/or
second-order turning points, in the considered spatial region, the presented
approach yields an analytical expression to be the regular approximate complex
solution to the one-dimensional Schr\"odinger equation.

\section{Acknowledgments}

The author expresses his gratitude to the Programm of supporting the leading
scientific schools of RF (grant No 2553.2008.2) for partial support of this work.


\section*{References}
\begin{thebibliography}{861}
\bibitem{Doo}
Maltsev N E 1994 {\it J Math Phys} {\bf 35} 1387

\bibitem{Ch10}
Chuprikov N L 1994 {\it Proceedings of International Simposium "Physics and
Engenering of Millimiter and Submillimiter Waves"} (Kharkov: Inst. of Radiophysics
and Electronics of National Academy of Science of Ukraine) 243

\bibitem{Ch11}
Chuprikov N L 1994 {\it Proceedings of International Simposium "Physics and
Engenering of Millimiter and Submillimiter Waves"} (Kharkov: Inst. of Radiophysics
and Electronics of National Academy of Science of Ukraine) 240

\bibitem{Ch12}
Chuprikov N L 1994 (Moscow: VINITI) B94 preprint (in Russian)

\bibitem{Adac}
Hyouguchi T, Adachi S, and Ueda M 2002 \PRL {\bf 88} 170404

\end{thebibliography}
\end{document}